\documentclass[twocolumn,showpacs,preprintnumbers,amsmath,amssymb]{revtex4}

\usepackage{graphicx}
\usepackage{graphics}
\usepackage{dcolumn}
\usepackage{bm}

\begin{document}

\title{Thermal expansion and compressibility of monogermanides of 3d-metals}

\author{G. A. Valkovskiy$^1$, E. V. Altynbaev$^{1, 2}$, M. D. Kuchugura$^2$, E. G. Yashina$^2$, V.~A.~Dyadkin$^{3}$, A.~V.~Tsvyashchenko$^4$, V.~A.~Sidorov$^4$, L.~N.~Fomicheva$^4$, M. Bykov$^5$, E. Bykova$^{5, 6}$, L. Dubrovinsky$^6$, D. Yu. Chernyshov$^7$, S. V. Grigoriev$^{1, 2}$}

\address{$^1$ Faculty of Physics, Saint-Petersburg State University, 198504 Saint-Petersburg, Russia\\
$^2$ Petersburg Nuclear Physics Institute, Gatchina, 188350 Saint-Petersburg, Russia\\
$^3$ European Synchrotron Radiation Facility, 38043 Grenoble, France\\
$^4$ Institute for High Pressure Physics, 142190 Troitsk, Moscow Region, Russia\\
$^5$ Laboratory of Crystallography, University of Bayreuth, 95440 Bayreuth, Germany\\
$^6$ Bavarian Research Institute of Experimental Geochemistry and Geophysics, University of Bayreuth, 95440 Bayreuth, Germany\\
$^7$ Swiss-Norwegian Beamlines at the ESRF, 38000 Grenoble, France}


\begin{abstract}
Synchrotron diffraction as a function of temperature and pressure, specific heat, magnetic susceptibility and small-angle neutron scattering experiments have revealed an anomalous response of MnGe. Similar but less pronounced behavior has also been observed in Mn$_{1-x}$Co$_x$Ge and Mn$_{1-x}$Fe$_x$Ge solid solutions. Spin density fluctuations and Mn spin state instability are discussed as possible candidates for the observed effects. 
\end{abstract}

\maketitle


\section{Introduction}

Cubic helimagnets with B20 structure have very simple crystal structure (P2$_1$3, two atoms only in the asymmetric part of $\approx$ 4\,\AA\,cubic unit cell) but surprisingly rich variety of magnetic phenomena. MnSi may serve as an example where complex chiral magnetic objects - spirals and skyrmions - inherit chirality from the crystal structure via the Dzyaloshinskiy-Moriya interaction \cite{Pfleiderer, Moskvin, GrigorievPRB2005, StishovPRB2007, GrigorievPRB2010}. 

MnGe, a compound isostructural to MnSi with a bigger unit cell dimension, shows even more complex magnetic response. First, in spite of larger Mn-Mn separation, the magnetic ordering temperature is much higher ($T^\mathrm{MnGe}_c \approx 170$ K and $T^\mathrm{MnSi}_c \approx 29$ K for MnSi) \cite{Altynbaev, DeutschBonville, DeutschMakarova}. Second, small angle neutron scattering revealed an intricate ordering scenario comprising new ferromagnetic phase between 150 and 300 K not met for MnSi \cite{Altynbaev}. Scattering data agree with inhomogeneous character of this phase that can be seen as a mixture of ferromagnetic droplets embedded in a paramagnetic matrix.

Provided that these materials are itinerant helimagnets, spin density fluctuations may affect fundamentally both their magnetic and thermodynamic properties \cite{Moriya, Takahashi}; this is a core of another coupling of crystal structure with magnetism - magneto-volume effect.  
Perhaps, the magneto-volume effect is the reason for MnSi to have the anomaly of a thermal expansion coefficient, firstly observed in \cite{Fawcett}. The authors of \cite{Matsunaga} have shown that the coefficient of thermal expansion exhibits a sharp drop at the magnetic ordering temperature along with a shoulder on the high-temperature side. An extensive study of MnSi thermal properties \cite{StishovPRB2007, StishovJPCM2008, StishovPRL} has uncovered similar shoulder for heat capacity on the high-temperature side of the corresponding peak at the phase transition; such pre-transition behaviour was tentatively linked to spin fluctuations.

Another instability, that may be related to the observed anomalies and potentially linked to the complexity of the observed magnetic response, is the spin state instability. It implies existence of two isostructural configurations of MnGe characterized by different Mn spin state and unit cell dimensions. Such spin state instability was recently proposed on the basis of {\it ab initio} calculations for MnGe \cite{Rossler, DiTusa}; neutron diffraction experiment as a function of temperature and pressure was interpreted in favor of this scenario \cite{DeutschMakarova}.

Here we report the results of synchrotron powder and single crystal diffraction experiments as a functions of temperature and pressure for MnGe  and for solid solutions  Mn$_{1-x}$Co$_x$Ge and Mn$_{1-x}$Fe$_x$Ge. The data on thermal expansion and compressibility are complemented with heat capacity, magnetization and small angle neutron scattering experiments. The derived results serve as a discriminative input for the discussion of microscopic reasons for non-conventional magnetic behavior of MnGe.

\section{Experimental}

The polycrystalline samples were synthesized under high pressure of 8 GPa by melting the constituent elements mixed in a stoichiometric ratio with an electric current. After, the samples were rapidly quenched to the room temperature where the pressure was released (see \cite{Tsvyashchenko1984} for details).

AC susceptibility measurements were done with a SR830 lock-in amplifier in a large compensated coil system (see \cite{Sidorov} for more details). 

Small-angle neutron scattering (SANS) experiments were carried out at SANS-1 at FRM-II (Munich, Germany). Neutrons with a mean wavelength of 0.6 nm were used. The sample-detector distance was set to~2.2~m. The scattering intensity was measured in zero-field cooling mode from 300 to 5 K.

Powder diffraction data were collected using Pilatus@SNBL diffractometer (Swiss Norwegian Beam Lines at ESRF, Grenoble, France). The beam size was set to 0.25 mm, a Pilatus 2M pixel area detector was used for data recording. Geometrical parameters of the diffractometer and the wavelength 0.6888\, \AA\, were calibrated with the LaB$_6$ NIST 660a standard. A sample was crushed in a fine powder and placed into 0.1 mm capillary. The temperature, controlled by Cryostream 700+, was varied in the 80 - 500 K range, with the 170 - 240 K/h rate and the 2.5 K step. The MnGe sample was measured additionally with the 1 K temperature step for a better temperature sampling. All the powder diffraction patterns were azimuthally integrated with FIT2D \cite{Hammersley1997}. In order to obtain temperature variation of the lattice parameter the Rietveld refinement was done for every temperature (sequential refinement) with the FULLPROF software package \cite{Rodriguez-Carvajal}.

High-pressure single-crystal x-ray diffraction experiments were carried out at ID09A (ESRF) using the monochromatic radiation with the wavelength of 0.41168~\, \AA\,. All the high pressure experiments were performed using a diamond anvil cell (DAC) technique. Pressure was generated by means of Le Toullec type DACs equipped with Boehler-Almax diamond anvils (0.25 mm culet sizes). The DACs were loaded with neon as pressure-transmitting medium. Pressure was determined using the ruby R1 fluorescence line as a pressure marker (up to 10 GPa) and using the equation of state of Ne (above 10 GPa) \cite{Fei}. X-ray diffraction images, taken upon continuous rotation of the DAC from -20$^\circ$ to +20$^\circ$ on omega (referred to as wide-scan images), were collected at every pressure step, and for several pressure points a complete data collection was performed by narrow 0.5$^\circ$ omega-scanning in the range from -32$^\circ$ to +32$^\circ$. All the high-pressure data were collected using a MAR555 flat panel detector. The treatment (integration and determination of the orientation matrix) was performed with the CrysAlisPro software \cite{Crysalis}. For the processing of the sets of intensities from wide-scan images and the refinement of the unit cell parameters GSE ADA  software \cite{ADA} was used.

Heat capacity measurements for MnGe were performed in a PPMS-9 (Quantum Design) working as a thermal-relaxation calorimeter. A micro-heater and thermometer are attached to the bottom side of the sample platform. The sample was mounted to the platform by using a thin layer of grease, which provided the required thermal contact to the platform. The vacuum greases Apiezon $N$ and Apiezon $H$ were used in the temperature ranges of 80 - 300 K and 200 - 400 K respectively. The grease heat capacity was considered as an instrumental function. The heat capacities of the sample measured with Apiezon $N$ and Apiezon $H$ in the range of 200 - 300 K are in a good agreement. Thorough inspection of the heat capacity was carried out by means of accurate subtraction of the grease contribution and small temperature step. In the following we deal with molar heat capacity, which is the heat capacity related to one mole of a substance.

\section{Results}

\subsection{Magnetic susceptibility and SANS}

\begin{figure}[h!]
\center{\includegraphics[width=1.05\linewidth]{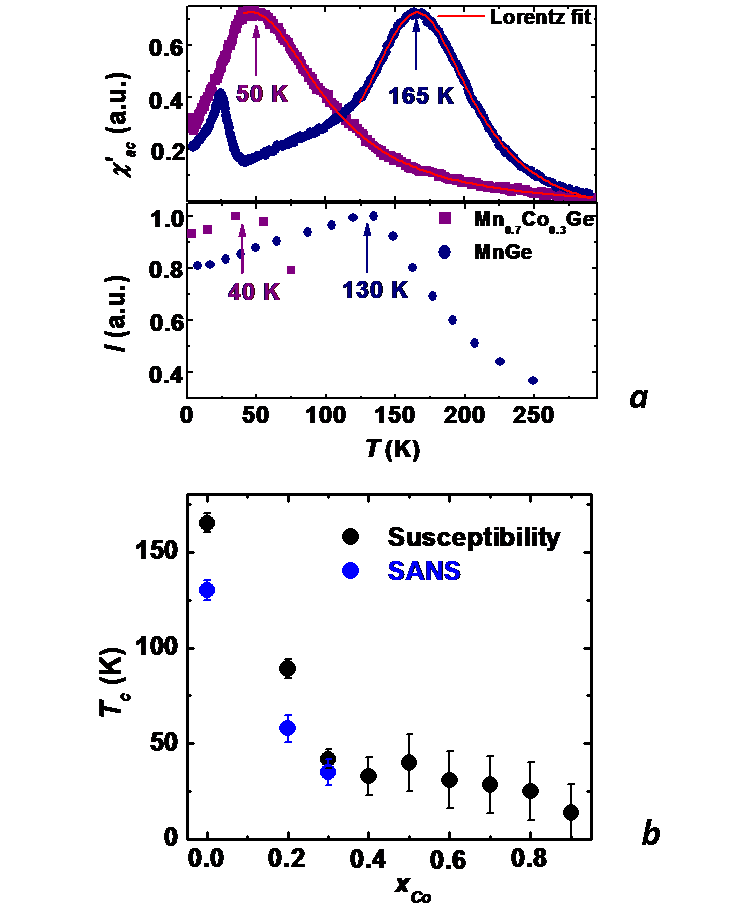}}
\caption{($a$) Upper panel: Temperature dependence of the real ac magnetic susceptibility $\chi_{ac}(T)$ for MnGe and Mn$_{0.7}$Co$_{0.3}$Ge. Lower panel: Temperature dependence of the SANS intensity of MnGe and Mn$_{0.7}$Co$_{0.3}$Ge integrated over the $Q$ range covered by the detector at zero field. ($b$) The characteristic temperatures related to the maxima of magnetic susceptibility and integrated SANS intensity.}
\label{fig:Fig1_SANS_susc}
\end{figure}

Fig. \ref{fig:Fig1_SANS_susc} $a$ (upper panel) shows  temperature dependence of the AC magnetic susceptibility for MnGe and Mn$_{0.7}$Co$_{0.3}$Ge. There is a broad peak at $\sim$~ 170 K in the curve for MnGe that is in a qualitative agreement with the previous findings \cite{Takizawa, Tsvyashchenko, Altynbaev, DeutschBonville, DeutschMakarova}. Similar broad asymmetric maxima are presented in the curves for the other samples under study. In order to get the position of maxima the curves are fitted with a Lorentzian.
Fig. \ref{fig:Fig1_SANS_susc} $a$ (lower panel) compares integrated intensity of small-angle neutron scattering for MnGe and Mn$_{0.7}$Co$_{0.3}$Ge; both SANS and susceptibility data are remarkably similar. Fig. \ref{fig:Fig1_SANS_susc} $b$ summarises the data on the characteristic temperatures for all the samples under study. The characteristic temperature as a function of cobalt concentration shows fast decrease with increasing cobalt concentration and saturates at $\approx$ 20 K for $x > 0.4$. 

\subsection{X-ray diffraction as a function of temperature}

The powder x-ray diffraction patterns for all the samples possess weak diffraction peaks of a few additional phases.
Therefore, only the unit cell dimensions are used for the thermal expansion evaluation. This approach enables us to probe predominantly the phase of interest \cite{Dyadkin}. 

\begin{figure}[h!]
\center{\includegraphics[width=1\linewidth]{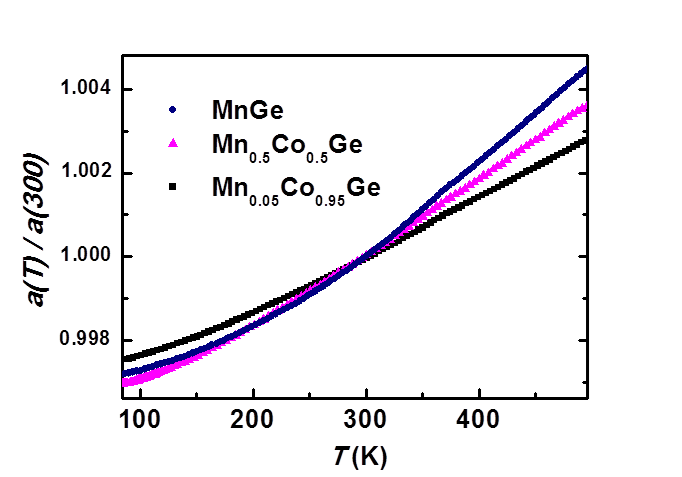}}
\caption{The temperature evolution of the unit cell parameters for Mn$_x$Co$_{1-x}$Ge normalized to the respective room temperature values, i.e. 4.7947 \, \AA\,, 4.7183 \, \AA\, and 4.6454 \, \AA\, for MnGe, Mn$_0.5$Co$_{0.5}$Ge and Mn$_0.05$Co$_{0.95}$Ge, respectively.}
\label{fig:Fig2_a_T}
\end{figure}

The temperature evolution of the unit cell dimensions for Mn$_x$Co$_{1-x}$Ge series is exemplified in Fig. \ref{fig:Fig2_a_T}. 
The dependences were normalized to the room temperature lattice parameters in order to highlight the difference in thermal expansion.

\begin{figure}[h!]
\center{\includegraphics[width=1\linewidth]{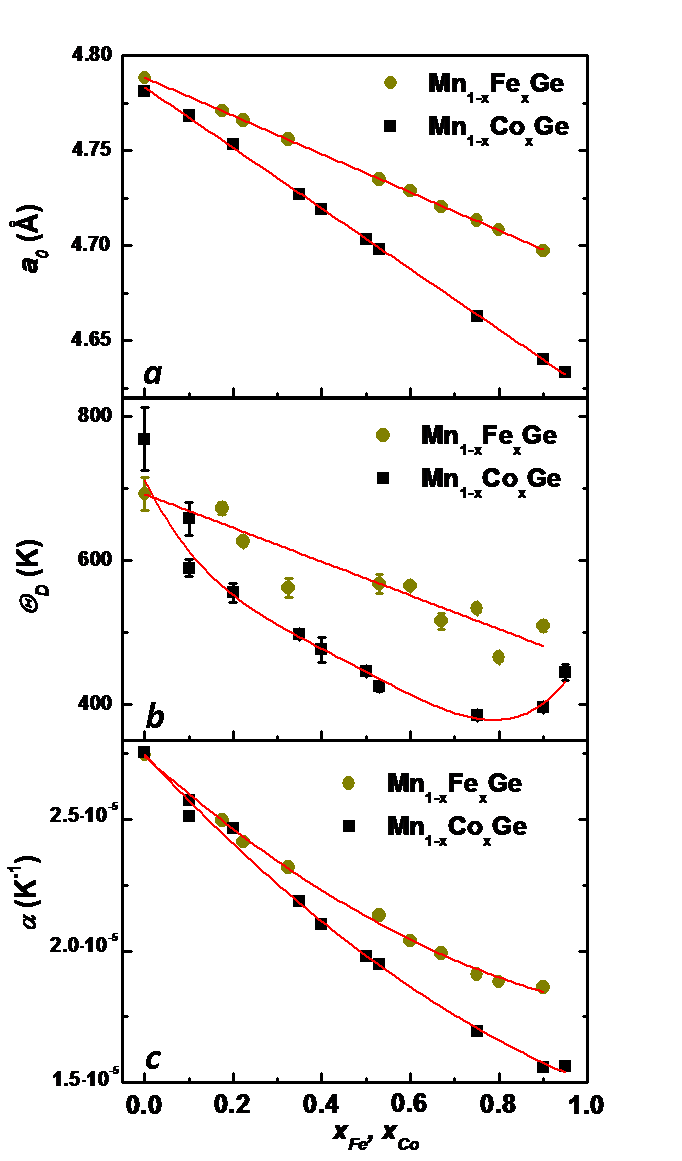}}
\caption{The concentration dependences obtained from the Debye-Gr{\"u}neisen parametrization of thermal expansion ($a$) the lattice parameter $a_0$ at $T$ = 0 K , ($b$) the Debye temperature $\Theta_D$  and ($c$) the high temperature asymptote of thermal expansion coefficient $\alpha$. The data are presented for Mn$_{1-x}$Fe$_{x}$Ge (circles) (from Ref. \cite{Dyadkin}) and for Mn$_{1-x}$Co$_{x}$Ge (squares). The solid curves are guides for the eyes.}
\label{fig:Fig3_a0ThetaDpref}
\end{figure}

Unit cell dimensions as a function of temperature have been parametrised with the Debye-Gr{\"u}neisen equation, see e.g. \cite{Sayetat, Dyadkin}.
The derived parameters are the following, $a_0$ is the unit cell dimension approximated to 0~K, $\Theta_D$ is the Debye temperature, $\alpha$ is the high temperature asymptote of thermal expansion coefficient. These parameters for Mn$_{1-x}$Fe$_{x}$Ge (from Ref. \cite{Dyadkin}) and for Mn$_{1-x}$Co$_{x}$Ge are shown in Figs. \ref{fig:Fig3_a0ThetaDpref} $a$ - $c$.

\begin{figure}[h!]
\center{\includegraphics[width=1\linewidth]{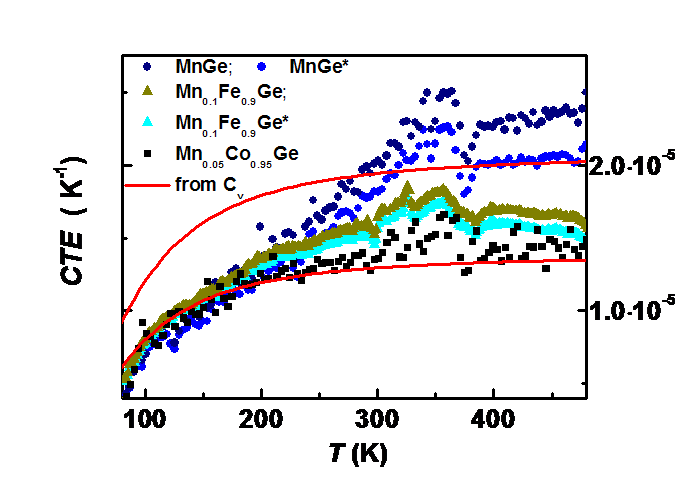}}
\caption{The temperature dependence of thermal expansion coefficient for the samples with high Mn (MnGe), Fe (Mn$_{0.1}$Fe$_{0.9}$Ge) and Co (Mn$_{0.05}$Co$_{0.95}$Ge) content. The data for MnGe and Mn$_{0.1}$Fe$_{0.9}$Ge minus the electronic contribution are marked with asterisks. The solid curves were obtained from simulated heat capacity data at constant volume in terms of Debye model with the Debye temperature of 350 K. The simulated heat capacity curves were multiplied by some scale factors. }
\label{fig:Fig4_CTE}
\end{figure}

The coefficient of thermal expansion ($CTE$) as a function of temperature have been calculated as 
\begin{equation}
 CTE = \frac{1} { a(T)} \frac{da(T)}{dT}
\end{equation}

and it is shown in Fig. \ref{fig:Fig4_CTE} for MnGe, Mn$_{0.1}$Fe$_{0.9}$Ge and Mn$_{0.05}$Fe$_{0.95}$Ge. It seen that at temperatures lower than $\sim 200$ K the $CTE$ for different materials are close, while for the temperatures above $\sim 200$ K the thermal expansion coefficient for MnGe is markedly enhanced.

\subsection{X-ray diffraction as a function of pressure}

\begin{figure}[h!]
\center{\includegraphics[width=1\linewidth]{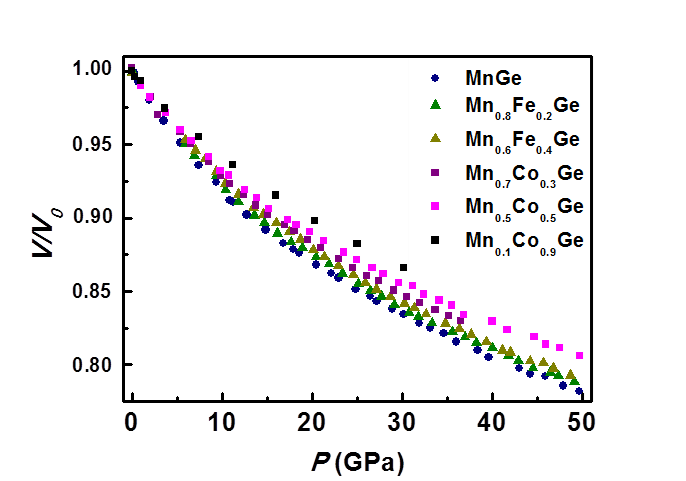}}
\caption{Equations of state for Mn$_{1-x}$Co$_{x}$Ge and Mn$_{1-x}$Fe$_{x}$Ge. }
\label{fig:Fig5_EOS}
\end{figure}

The experimental dependences of the unit cell volume on pressure (equation of state, EOS) are shown in Fig. \ref{fig:Fig5_EOS}. A monotonic variation with increasing Co or Fe content is evident. 

\begin{figure}[h!]
\center{\includegraphics[width=1\linewidth]{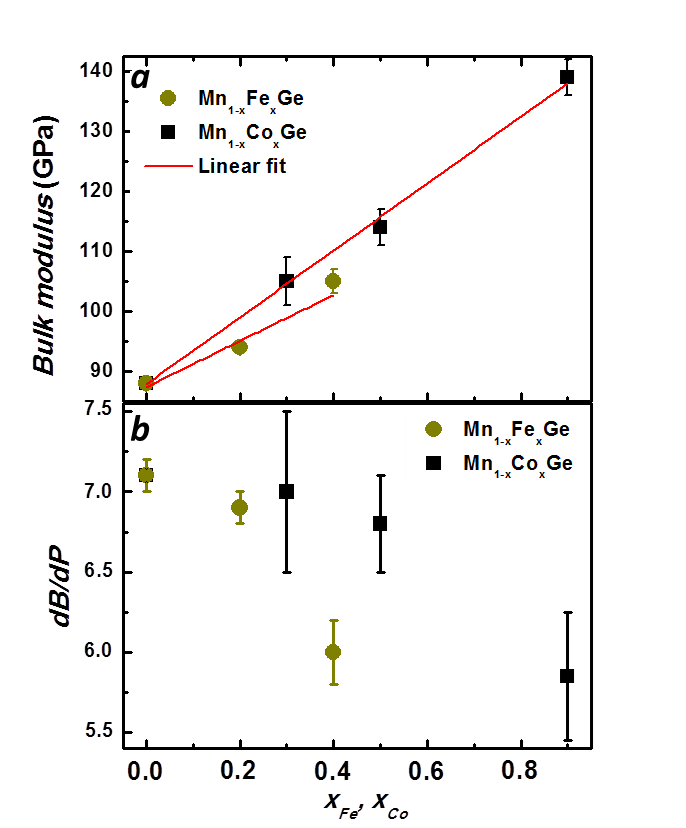}}
\caption{The concentration dependence of the bulk modulus ($a$) and the bulk modulus pressure derivative ($b$) for Mn$_{1-x}$Fe$_{x}$Ge (circles) and Mn$_{1-x}$Co$_{x}$Ge (squares) obtained from EOS, solid lines represent linear fits.  }
\label{fig:Fig6_B_dBdP}
\end{figure}

The EOS have been parametrised using the 3rd order Birch-Murnaghan equation \cite{Birch}. The obtained parameters, i.e. the bulk modulus and its pressure derivative are given in Fig. \ref{fig:Fig6_B_dBdP} $a$ and Fig. \ref{fig:Fig6_B_dBdP} $b$, respectively. One can see that the bulk modulus increases and its pressure dependence decreases with Mn substitution with Fe or Co. The behavior of the bulk modulus has been well parametrised with a linear function. The pressure derivative of the bulk modulus is turned out to be enhanced for MnGe. This quantity is likely to vary weaker in the range of low cobalt concentrations than for high cobalt concentrations. 

\begin{figure}[h!]
\center{\includegraphics[width=1\linewidth]{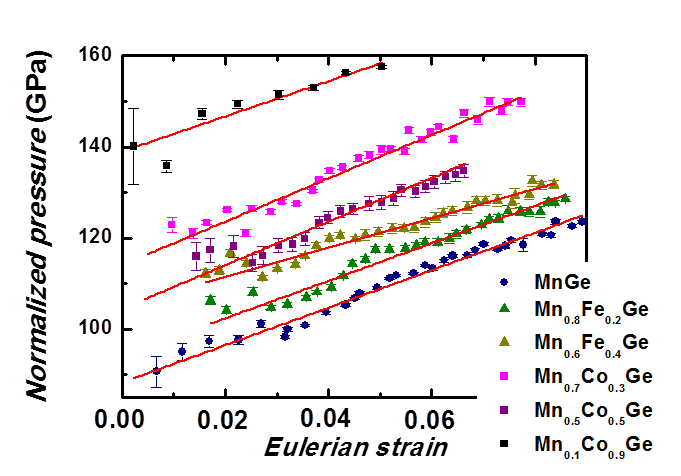}}
\caption{Ff - plots for Mn$_{1-x}$Co$_{x}$Ge and Mn$_{1-x}$Fe$_{x}$Ge - experimental (points), and fitted (solid lines).  }
\label{fig:Fig7_Ff}
\end{figure}

In order to examine the scenario based on spin state transition as proposed in Ref. \cite{DeutschMakarova}  we used the F-f plots (Eulerian stress-strain diagram, see \cite{Angel}) presented in Fig. \ref{fig:Fig7_Ff}. All the dependences are essentially linear within the experimental accuracy. That implies absence of any abrupt transitions at room temperature in the studied pressure range.

\subsection{Heat Capacity}

\begin{figure}[h!]
\center{\includegraphics[width=1\linewidth]{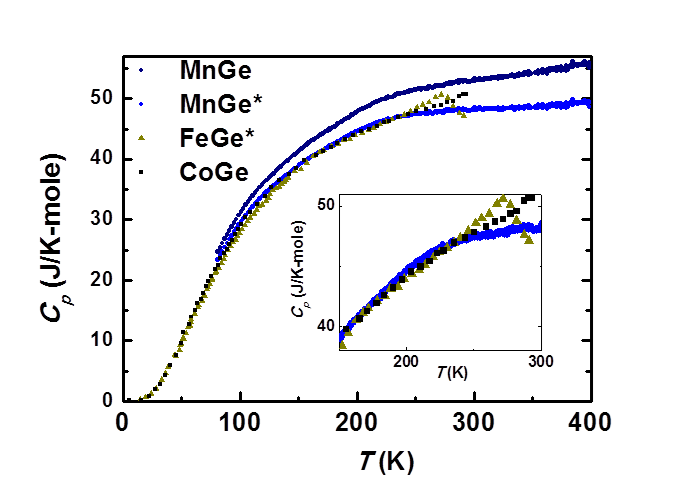}}
\caption{The experimental heat capacity. The data for FeGe, CoGe and low-temperature data for MnGe are from Refs. \cite{DiTusa} and \cite{Tsvyashchenko}.  The data for MnGe and FeGe minus the electronic contribution are marked with asterisks. The inset compares the heat capacities without electronic contribution more closely.}
\label{fig:Fig8_HC}
\end{figure}

The experimental heat capacity at constant pressure $C_p$ for MnGe is presented in Fig. \ref{fig:Fig8_HC}. Fig. \ref{fig:Fig8_HC} also shows literature data from Ref. \cite{DiTusa} on  heat capacity of CoGe, which does contain only phonon contribution. In addition, the heat capacities after electronic contribution subtraction for MnGe and FeGe are given. In the following we will consider these heat capacities without electronic contribution.   

At variance with FeGe no extra peak at Curie temperature is observed for MnGe, thus any first order transition can be ruled out \cite{DiTusa}. The heat capacity data for MnGe  systematically deviates from that for non-magnetic CoGe on heating, see Fig. \ref{fig:Fig8_HC} and Fig. \ref{fig:Fig9_HC_magn}.

\section{Discussion and conclusion}

Fig. \ref{fig:Fig3_a0ThetaDpref} ($a$ and $c$) shows that the concentration dependences of $a_0$ and $\alpha$ (obtained from the Debye-Gr{\"u}neisen parametrisation of the thermal expansion data) are quite linear. It is seen in Fig. \ref{fig:Fig3_a0ThetaDpref} $b$ that the Debye temperatures obtained from the parametrization varies in the broad range from 450 K for Mn$_{0.05}$Co$_{0.95}$Ge to about 750 K for MnGe. However, recent measurements of the heat capacities of MnGe, FeGe and CoGe \cite{DiTusa} correspond to characteristic Debye temperatures much lower (of about 300~K) than those fitted from the temperature dependences of the unit cell dimension. Besides, Debye temperature for the structurally similar compounds is unlikely to differ by hundreds of Kelvin. For example, substitution of Mn by Fe or Co in MnSi keep Debye temperature unchanged \cite{Bauer}.

Modelling of the heat capacity for CoGe within the Debye model resulted in the Debye temperature of 350 K. The estimation of Debye temperature from bulk modulus using Moruzzi scaling factor \cite{Moruzzi} resulted in close value, i.e.  $\sim 320$ K. Fitting of the heat capacity for MnGe with excluded electronic contribution resulted in similar Debye temperature as for CoGe $\sim 350$ K. So, the comparison of the Debye temperatures from the heat capacities and $CTE$ data suggests that the fitting of the $CTE$ with the Debye-Gr{\"u}neisen model represents a certain parametrization with some effective parameters.   

According to the Gr{\"u}neisen quasi-harmonic model the temperature dependence of heat capacity at constant volume and $CTE$ should show the same temperature dependence (see e.g. \cite{Grimvall}). We have multiplied the Debye heat capacity with the Debye temperature of 350 K at a certain scaling factor in order to match the heat capacity data with the $CTE$ data. One can obtain a good agreement in the case of Mn$_{0.05}$Co$_{0.95}$Ge (Fig. \ref{fig:Fig4_CTE}). The corresponding Gr{\"u}neisen parameter derived from the scaling factor is 1.9. Moreover, fitting the $CTE$ for this sample with fixed Debye temperature of 350 K resulted in fairly good agreement with the same Gr{\"u}neisen parameter. For MnGe this approach works either above room temperature or below 200 K. The temperature evolution of $CTE$ may be seen as a crossover between two states limited by the solid curves in Fig. \ref{fig:Fig4_CTE}. These two states can be associated with different spin configurations linked to different unit cell volumes and effective magnetic moments per formula unit. The evolution of $CTE$ on cooling indicates a gradual change in the average fraction of high spin state. However, according to susceptibility and neutron scattering data the conversion is not complete even at very low temperatures \cite{DiTusa, DeutschBonville, DeutschMakarova}. Such a scenario assumes that ferromagnetic droplets observed with neutron scattering should be associated with regions enriched with high spin states.

\begin{figure}[h!]
\center{\includegraphics[width=1\linewidth]{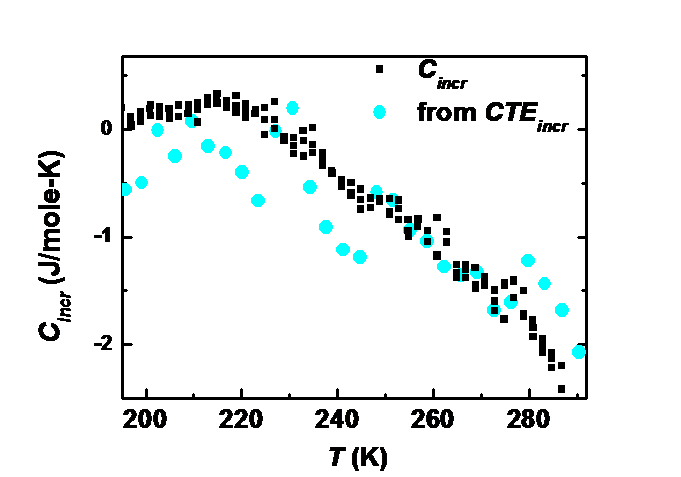}}
\caption{The heat capacity difference between MnGe (without electronic contribution) and CoGe. The heat capacity increment is compared with the thermal expansion increment (the difference between MnGe and Mn$_{0.05}$Co$_{0.95}$Ge thermal expansion coefficient) multiplied by a negative scale factor.}
\label{fig:Fig9_HC_magn}
\end{figure}

The other possible interpretation of the observed peculiarities for MnGe is a magneto-volume effect linked to spin fluctuations. Comparing the heat capacities for MnGe (without electronic contribution) with that of non-magnetic CoGe (inset in Fig. \ref{fig:Fig8_HC}) one sees a negative contribution for MnGe above the ordering temperature. This anomaly progressively evolves with temperature. Notably, this incremental contribution to the heat capacity of MnGe over that of CoGe can be scaled with the corresponding $CTE$ data (Fig. \ref{fig:Fig9_HC_magn}), herewith the scaling factor turned out to be negative. It is worth noting that negative magnetic heat capacity has also been observed for MnSi at the temperatures higher the helimagnetic ordering temperature and tentatively linked to spin fluctuations \cite{StishovPRL}. Such a scenario assumes strong fluctuations of spin density within the temperature range of existence of ferromagnetic droplets and agrees with absence of the effect in non-magnetic CoGe. It states however unclear why no trace of such a response has been observed for the thermal expansion, heat capacity, magnetic susceptibility and neutron scattering of isostructural but magnetically ordered FeGe.

The diffraction experiments as a function of pressure at room temperature show no sharp transition which can be associated with abrupt change of spin state for all compounds. However, unusually high value of the bulk modulus pressure derivative for Mn-rich materials may indicate a gradual crossover towards lower volume low spin state that is more favorable at high pressures.

Remarkably, at room temperature MnGe has not only enhanced $CTE$, but also a reduced bulk modulus as compared with isostructural Mn$_{1-x}$Co$_{x}$Ge and Mn$_{1-x}$Fe$_{x}$Ge materials. The obtained bulk modulus $\approx$ 88~GPa is lower than calculated ones (133~GPa \cite{Rossler}; 238~GPa \cite{DiTusa}) since calculations do not account for temperature and pressure induced change of the spin state. Moreover, our experimental value is lower than that reported in \cite{DeutschMakarova} - 106 GPa. The apparent difference may be linked to different pressure range (0 - 50 GPa in our work) and (0-10 GPa \cite{DeutschMakarova}). Also, the bulk modulus pressure derivative was fixed at value 5 in \cite{DeutschMakarova} and refined in our analysis.

To conclude, we have observed and characterized an anomalous thermodynamic response on the complex magnetic ordering scenario in MnGe. The non-phonon contributions to the thermal expansion, compressibility, and heat capacity persist up to high temperatures. They manifest themselves by enhanced value of thermal expansion coefficient, reduced bulk modulus, and unusually high value of the bulk modulus pressure derivative. In the same temperature range an inhomogeneous magnetic phase has been revealed by neutron diffraction; we believe that it is not fortunate coincidence but illustrate a link between macroscopic properties and some kind of correlated disorder. We propose spin state instability as the most probable candidate of such a disorder.

\section{Acknowledgements}

The authors acknowledge SNBL and ID09A beamlines of ESRF for their hospitality. The work is supported in part by Saint-Petersburg State University and by the RFBR (projects N$^\circ$ 14-22-01113 and 14-02-00001). The Interdisciplinary Resource Center of Functional Material Diagnostics for Medicine, Pharmocology and Nanoelectronics of St. Petersburg State University is acknowledged for the heat capacity measurements. The Centre for X-ray Diffraction Studies of St. Petersburg State University is acknowledged for laboratory x-ray diffraction studies.

\end{document}